\documentclass[aps,prb,twocolumn,superscriptaddress]{revtex4}
\usepackage{graphicx}
\usepackage{bm}
\usepackage{epsfig} 
\usepackage{amsfonts}
\usepackage{amsmath}

\allowdisplaybreaks[1] 

\newcommand*{\be}{\begin{equation}}
\newcommand*{\ee}{\end{equation}}
\newcommand*{\bea}{\begin{eqnarray}}
\newcommand*{\eea}{\end{eqnarray}}

\newcommand*{\sd}{^{\dagger}}

\def\bra#1{\langle #1|}
\def\ket#1{|#1\rangle}

\begin{document}

\title{The short-range correlations of a doped Mott insulator}

\author{Tiago C. Ribeiro}
\affiliation{Department of Physics, University of California, Berkeley, California 94720, USA}
\affiliation{Material Sciences Division, Lawrence Berkeley National Laboratory, Berkeley, California 94720, USA}

\date{\today}

\begin{abstract}
This paper presents numerical studies of the single hole $tt't''J$ model 
that address the interplay between the kinetic energy of itinerant 
electrons and the exchange energy of local moments as of interest 
to doped Mott insulators.
Due to this interplay, two different spin correlations coexist around
a mobile vacancy.
These \textit{local} correlations provide an effective two-band picture
that explains the two-band structure observed in various theoretical and 
experimental studies, the doping dependence of the momentum space
anisotropic pseudogap phenomena and the asymmetry between hole and 
electron doped cuprates.
\end{abstract}


\maketitle

\section{\label{sec:intro}Introduction} 

The evolution between the weakly correlated Fermi metal and the strongly 
coupled Mott insulator is a major and long-standing problem in the
field of condensed matter physics. \cite{M4916, A5902, H6338}
It concerns a vast list of material compounds \cite{IF9839} where the 
local character of $d$ and $f$ orbitals enhances the electron effective 
mass together with the role of the electron-electron Coulomb repulsive 
interaction.
The resulting competition between the small kinetic energy and the 
strong interaction may lead itinerant electrons in the metallic
state to form local moments in the Mott insulating state. \cite{M4916}
In this paper, I focus on the interplay between such itinerant and 
localized electrons.

The generalized-$tJ$ model explicitly embodies the above interplay.
Indeed, ``$t$'' stands for the kinetic energy term of itinerant charge 
carriers and ``$J$'' stands for the interaction term between localized 
spins.
The intricacy of this model follows from the mutual frustration 
between these two terms. 
Specifically, the $J > 0$ term favors a staggered moment spin background 
that constrains the motion of vacancies, while the $t$ term moves electrons
around and, thus, reshuffles and destroys the underlying 
antiferromagnetic (AF) spin pattern.
The two-dimensional (2D) $tt't''J$ model, which this paper addresses, 
is especially interesting because the compromise between the 
spin exchange and hole kinetic energies is particularly subtle in 
the parameter regime of interest to real materials, such as the 
high-temperature superconducting cuprates. \cite{A8796,D9463,LN0617}

Following the recent improvement in computational resources, experimental 
resolution and sample quality, various non-trivial results have illuminated 
our understanding of 2D doped Mott insulators.
It is exciting to note that many of these results are consistently
obtained by different theoretical approaches and by experiments.
For instance, a variety of numerical and analytical studies show that
the electronic spectrum below the chemical potential has a robust
two-band structure and that changing the electron density 
redistributes the spectral weight in a momentum dependent way.
\cite{PH9544,MH9545,DZ0016,GE0036,TS0009,T0417,RW0501,KK0614,PB0402,CC0502}
Similar conclusions apply to the two dispersive features displayed by
angle-resolved photoemission spectroscopy (ARPES) on the cuprates.
\cite{RS0301,KS0318,YZ0301,SR0402,DS0373}
Since the electron dynamics in strongly correlated systems follows
from the local environment around the carriers, \cite{SHORTRANGE}
the above two-band structure reflects the presence of two local 
correlations, which arise due to the interplay between itinerant 
electrons and local moments. 

This paper explores the microscopic origin of the above short-range
correlations and, thus, of the aforementioned
two-band structure that appears in both theory and experiments.
Specifically, in Sec. \ref{sec:local} the exact diagonalization and 
the self-consistent Born approximation techniques are employed
to study the single hole problem in the $tt't''J$ model.
I show that the interplay between the ``$t$'' and ``$J$'' terms 
of the Hamiltonian translates into the coexistence of two different types 
of spin correlations around the vacancy --
one type is driven by the kinetic energy term and the other by the 
exchange energy term.
These short-range correlations, which follow from purely
local energetic considerations and whose properties are studied in
Sec. \ref{sec:properties}, underlie a diverse set of non-trivial 
and, by now, well established properties of 2D doped Mott 
insulators.
These include the doping dependence of the pseudogap dispersion and of 
the pseudogap momentum space spectral weight distribution, 
\cite{DS0373,ZY0401,YZ0301,RS0301,KF0517,TY0403,AR0201} as 
well as the asymmetry between the hole and electron doped regimes of the 
cuprate compounds (Sec. \ref{sec:anisotropy}). 
\cite{DS0373,KW9845,YZ0301,RS0301,AR0201,TM9496,T0417,RW0501,KK0614}

\section{\label{sec:local}Two local correlations}

\subsection{\label{subsec:model}The model system}

The single hole 2D $tt't''J$ Hamiltonian is
\be
H_{tt't''J} = - \sum_{\langle ij \rangle, \sigma} t_{ij} \left(
\widetilde{c}_{i,\sigma}\sd \widetilde{c}_{j,\sigma} + H.c.\right) +
\sum_{\langle ij \rangle} J_{ij} \bm{S}_i.\bm{S}_j
\label{eq:Htj}
\ee
where $\widetilde{c}_{i,\sigma}$ is the constrained electron 
operator $\widetilde{c}_{i,\sigma} = c_{i,\sigma} (1-n_{i,-\sigma})$.
$t_{ij}$ equals $t$, $t'$ and $t''$ for first, second and third 
nearest-neighbor (NN) sites respectively and vanishes otherwise.
The exchange interaction only involves NN spins for 
which $J_{ij}=J$.
In this paper $J \in [0.2 , 0.8]$ (units are set so that $t=1$),
which includes the experimentally relevant regime $J\approx0.4$.
The calculations are not extended down to $J=0$ because, in that limit,
the hole is subjected to the Nagaoka instability \cite{N6692} and, thus,
the physics for $J\approx0$ is specific to such a regime and is not 
relevant to materials like the cuprates. \cite{WA0111}
The calculations for $J>0.8$ do not change the argument below nor
the consequent conclusions.

As mentioned in Sec. \ref{sec:intro}, this paper addresses the
electron dynamics as probed by the electron spectral function.
Its focus does not lie in the full details of the
spectral function line shape, but rather on the fact that the 
electronic spectrum displays two separate dispersive features below 
the Fermi level.
This work also concerns the momentum distribution of electron 
spectral weight, which displays distinct behavior in separate 
regions of the Brillouin zone, namely the regions around 
$(\pi/2,\pi/2)$ and $(\pi,0)$.
The above facts encode short time and short length scale physics.
Hence, within the above context, it is relevant to study 
small lattice systems and,
unless otherwise stated, all the results below come from the exact 
diagonalization of $H_{tt't''J}$ on a $4 \times 4$ lattice.
The exact diagonalization analysis is further substantiated by results
from the self-consistent Born approximation approach to the 
spinless-fermion Schwinger-boson representation of the $tJ$ model 
\cite{RV8893,MH9117,RH9808} on a $16 \times 16$ lattice.

\subsection{\label{subsec:onehole}One-hole states}

There exist two extreme limits where the interplay between 
itinerant electrons and local moments occurs, namely the one where:
\textit{(i)} most electrons are itinerant and the corresponding
Fermi energy is the highest energy scale in the problem;
\textit{(ii)} most electrons form local moments and the system
reduces to a lattice of spins with a few mobile vacancies.
The former case is captured by the well understood Kondo model, which
addresses how the Fermi sea accommodates the presence of a local moment.
\cite{W7573} 
The second case, which is of interest close to the Mott insulator 
transition, differs from the standard Kondo lattice problem since the 
spin-spin interaction is larger than the itinerant electrons' Fermi 
energy. \cite{RW0674}
In this case, it is rather convenient to consider how the spin background 
adjusts to the presence of a hole. 

Hence, in what follows, one studies the lowest energy configurations 
of the spin background around a single vacancy.
In particular, one considers the lowest energy single hole state
$\ket{\psi_{\bm{k}},J,t't,''}$ for each momentum $\bm{k}$, where
$J$, $t'$ and $t''$ label the model parameters that define
the corresponding Hamiltonian $H_{tt't''J}$. 
This state can fall into two categories -- it either has zero
or non-zero quasiparticle spectral weight
$\left| \bra{\psi_{\bm{k}},J,t',t''} \widetilde{c}_{\bm{k},\sigma} 
\ket{\text{HF GS}} \right|^2$, where $\ket{\text{HF GS}}$ denotes the 
groundstate of the half-filled system.
For all $\bm{k}$, $t'$ and $t''$ there exists a certain
$J_c(\bm{k},t',t'')$ such that $\ket{\psi_{\bm{k}},J,t',t''}$
has zero quasiparticle spectral weight if and only if 
$J \leq J_c(\bm{k},t',t'')$.
The intuition behind this result is that for $J/t \gg 1$ the large
spin stiffness renders the spin background robust to the hole motion,
while for small enough $J/t$ the soft AF spin configuration is 
dramatically modified by the doped hole
(the Nagaoka instability \cite{N6692} perfectly illustrates this last 
case).
If $J \leq J_c(\bm{k},t',t'')$ one denotes 
$\ket{\psi_{\bm{k}},J,t't,''}$ by $\ket{\widetilde{U}_{\bm{k}},J,t',t''}$
(hence, by definition, $\ket{\widetilde{U}_{\bm{k}},J,t',t''}$
has vanishing quasiparticle spectral weight).
If, instead, $J > J_c(\bm{k},t',t'')$ the single hole state
$\ket{\psi_{\bm{k}},J,t't,''}$ can be approximately recast
as
\begin{align}
\ket{\psi_{\bm{k}},J,t',t''} \ &\cong \ q(\bm{k},J,t',t'') 
\ket{Q_{\bm{k}},t',t''} + \notag \\
& \quad \quad + u(\bm{k},J,t',t'') \ket{U_{\bm{k}},t',t''}
\label{eq:decomp}
\end{align}
where $\ket{Q_{\bm{k}},t',t''}$ and $\ket{U_{\bm{k}},t',t''}$ 
are orthonormal states (to be defined below) that do \textit{not} 
depend on $J$, \cite{Utilde}
while $q(\bm{k},J,t',t'')$ and $u(\bm{k},J,t',t'')$
are $J$-dependent coefficients that obey the normalization condition
$\left|q(\bm{k},J,t',t'')\right|^2 +
\left|u(\bm{k},J,t',t'')\right|^2 = 1$.
Eq. \eqref{eq:decomp}, which applies in a large range of 
$J$ values, is a major result in this paper.
It implies that, in a large interval of values of 
$J > J_c(\bm{k},t',t'')$, the eigenstates 
$\ket{\psi_{\bm{k}},J,t't,''}$ define a line parameterized by $J$ 
which approximately lies in a 2D plane in the single hole $tt't''J$ 
model Hilbert space.
The physical content of this statement, 
together with evidence supporting Eq. \eqref{eq:decomp}, are 
presented below.

\begin{table}
\begin{ruledtabular}
\begin{tabular*}{\hsize}{cc|cccccc}
& J & 0.3 & 0.4 & 0.5 & 0.6 & 0.7 & 0.8 \\
\hline
& $(\tfrac{\pi}{2},\tfrac{\pi}{2})$ & 0.9994 & 0.9994 & 0.9998 & 1 & 0.9998 & 0.9990 \\ 
$t'=0$ & $(\pi,0)$ & 0.9994 & 0.9994 & 0.9998 & 1 & 0.9998 & 0.9990 \\
$t''=0$ & $(\pi,\tfrac{\pi}{2})$ & 0.9972 & 0.9977 & 0.9993 & 1 & 0.9992 & 0.9970 \\
ED & $(\tfrac{\pi}{2},0)$ & 0.9975 & 0.9980 & 0.9994 & 1 & 0.9994 & 0.9977 \\
& $(0,0)$ & 0.9946 & 0.9923 & 0.9963 & 1 & 0.9938 & 0.9766 \\ 
\hline
& $(\tfrac{\pi}{2},\tfrac{\pi}{2})$ & 0.9996 & 0.9996 & 0.9998 & 1 & 0.9998 & 0.9990 \\ 
$t'=0$ & $(\pi,0)$ & 0.9994 & 0.9994 & 0.9998 & 1 & 0.9997 & 0.9986 \\ 
$t''=0$ & $(\pi,\tfrac{\pi}{2})$ & 0.9989 & 0.9988 & 0.9996 & 1 & 0.9995 & 0.9978 \\
SCBA & $(\tfrac{\pi}{2},0)$ & 0.9989 & 0.9988 & 0.9996 & 1 & 0.9995 & 0.9978 \\
& $(0,0)$ & 0.9016 & 0.9005 & 0.9766 & 1 & 0.9842 & 0.9488 \\ 
\hline
 & $(\tfrac{\pi}{2},\tfrac{\pi}{2})$ & 0.9994 & 0.9994 & 0.9998 & 1 & 0.9998 & 0.9990 \\ 
$t'=-0.2$ & $(\pi,0)$ & -- & 1  & 0.9998 & 0.9997 & 0.9999 & 1 \\
$t''=0.1$ & $(\pi,\tfrac{\pi}{2})$ & 0.9936 & 0.9952 & 0.9986 & 1 & 0.9987 & 0.9950 \\
ED & $(\tfrac{\pi}{2},0)$ & 0.9907 & 0.9943 & 0.9986 & 1 & 0.9988 & 0.9957 \\
& $(0,0)$ & 0.9880 & 0.9856 & 0.9943 & 1 & 0.9940 & 0.9807 \\ 
\end{tabular*}
\end{ruledtabular}
\caption
{\label{tab:check2d} Square of the overlap of 
$\ket{\psi_{\bm{k}},J,t',t''}$ with the Hilbert space 
$\{ \ket{\psi_{\bm{k}},J=0.2,t',t''}, \ket{\psi_{\bm{k}},J=0.6,t',t''} \}$
for different $J$ and $\bm{k}$. 
Both exact diagonalization (ED) and self-consistent Born approximation (SCBA) 
results are shown for $t',t''=0$.
Exact diagonalization results are also shown for $t'=-0.2$, $t''=0.1$.
For $t'=-0.2$, $t''=0.1$ and $\bm{k}=(\pi,0)$ the Hilbert space 
$\{\ket{\psi_{\bm{k}},J=0.4,t',t''}, \ket{\psi_{\bm{k}},J=0.8,t',t''} \}$
is used instead.}
\end{table}

If $t',t''=0$ then $J_c(\bm{k},t'=0,t''=0) < 0.2$ for all $\bm{k}$ in 
Table \ref{tab:check2d}.
This table shows that for all $J \in [0.2 , 0.8]$, as well as 
for all depicted momenta $\bm{k}$,
the states $\ket{\psi_{\bm{k}},J,t'=0,t''=0}$
have almost unit overlap with the 2D Hilbert space
$\{\ket{\psi_{\bm{k}},J=0.2,t=0',t''=0}, 
\ket{\psi_{\bm{k}},J=0.6,t'=0,t''=0} \}$.
This conclusion is further substantiated by the self-consistent Born 
approximation technique on a $16\times 16$ lattice (see
Table \ref{tab:check2d}), thus showing that 
the above result is not specific to the $4\times 4$ lattice used in 
the exact diagonalization calculation. \cite{CHOICE}
A very similar observation holds when $t',t'' \neq 0$, as 
Table \ref{tab:check2d} illustrates for $t'=-0.2$, $t''=0.1$.
The only difference between the above $t',t''=0$ and $t'=-0.2$, $t''=0.1$
cases is that $J_c(\bm{k}=(\pi,0),t'=0, t''=0) < 0.2$
while $0.3 < J_c(\bm{k}=(\pi,0),t'=-0.2, t''=0.1) < 0.4$.
Since the approximate equality in Eq. \eqref{eq:decomp} only applies for
$J > J_c(\bm{k},t',t'')$, in Table \ref{tab:check2d} one uses the 2D 
Hilbert space $\{\ket{\psi_{\bm{k}=(\pi,0)},J=0.4,t'=-0.2, t''=0.1}, 
\ket{\psi_{\bm{k}=(\pi,0)},J=0.8,t'=-0.2, t''=0.1} \}$
to illustrate that Eq. \eqref{eq:decomp} also applies when 
$\bm{k}=(\pi,0)$ and $t'=-0.2$, $t''=0.1$.

The above numerical results show that Eq. \eqref{eq:decomp} 
is a very good approximation for a wide range of 
values of the exchange coupling $J$. \cite{WIDERANGE}
However, one is still free to choose any orthonormal pair of states 
$\ket{Q_{\bm{k}},t',t''}$ and $\ket{U_{\bm{k}},t',t''}$
in the 2D Hilbert space used as a reference.
A physically sensible choice comes from requiring
$q(\bm{k},J,t',t'')$ to monotonously increase with $J/t$  
[thus $u(\bm{k},J,t',t'')$ monotonously decreases with $J/t$].
Since cranking up $J$ enhances the quasiparticle features of doped 
carriers, \cite{MH9117} the above condition is automatically satisfied 
if $\ket{U_{\bm{k}},t',t''}$ has vanishing quasiparticle spectral weight. 
This prescription uniquely determines Q states 
($\ket{Q_{\bm{k}},t',t''}$) and U states ($\ket{U_{\bm{k}},t',t''}$)
which, one should note, are not eigenstates 
of $H_{tt't''J}$. \cite{SINGLEJ}
The above construction implies that Q states bear the electron-like
properties of the true eigenstates $\ket{\psi_{\bm{k}},J,t't,''}$ and,
indeed, for all values of $t'$ and $t''$ used throughout this paper
one has that
$0.5 \lesssim
\frac{|\bra{Q_{\bm{k}}} \widetilde{c}_{\bm{k},\sigma} \ket{\text{HF GS}}|^2}
{|\bra{\text{HF GS}} \widetilde{c}_{\bm{k},\sigma}\sd 
\widetilde{c}_{\bm{k},\sigma} \ket{\text{HF GS}}|} \lesssim 0.8$
in the momentum space region around the $(\pi,0)-(0,\pi)$ 
line. \cite{MBZB}

\begin{figure}
\includegraphics[width=0.48\textwidth]{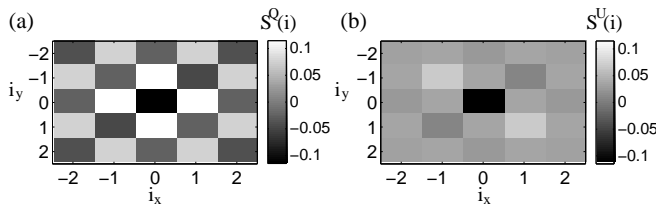}
\caption{\label{fig:spin}
(a) $S_{\bm{k}}^Q(\bm{i})$ and (b) $S_{\bm{k}}^U(\bm{i})$ 
where $\bm{i}$ is the distance to the vacancy (black square at the 
center), $\bm{k}=(\tfrac{\pi}{2},\tfrac{\pi}{2})$ and $t',t''=0$.
Different $\bm{k}$, $t'$ and $t''$ lead to qualitatively similar 
conclusions, as seen in Fig. \ref{fig:staggered}.
}
\end{figure}

The previous argument clarifies how the spin background adjusts to 
the presence of a moving hole.
Specifically, spins show two different types of correlations -- one 
type is enhanced upon increasing $J/t$ and the other becomes more 
pronounced when $J/t$ is reduced.
By definition, Q and U states capture these correlations and,
not surprisingly, they display distinct physical properties.
Simply based on the above energetic considerations, one expects
the former states to retain the AF correlations of the 
undoped system, while the doping induced spin correlations in U states 
facilitate hole hopping.
The analysis in Sec. \ref{subsec:r_and_k} confirms
this microscopic picture.
In principle, a similar construction 
applies to models other than the 2D $tt't''J$ model.
The significant fact about this model is that, for experimentally 
relevant parameters, the overlap of both Q and U states with 
$\ket{\psi_{\bm{k}},J,t',t''}$ is large and exhibits a 
considerable momentum dependence (see Sec. \ref{sec:anisotropy}). \cite{3D}

\section{\label{sec:properties}Properties of the local correlations}

The construction in Sec. \ref{subsec:onehole} identifies two
different spin configurations that coexist around a mobile
vacancy.
It also provides a recipe to separately obtain these
configurations and, thus, to study their properties.

\subsection{\label{subsec:r_and_k}Real and momentum space properties}

\begin{figure}
\includegraphics[width=0.48\textwidth]{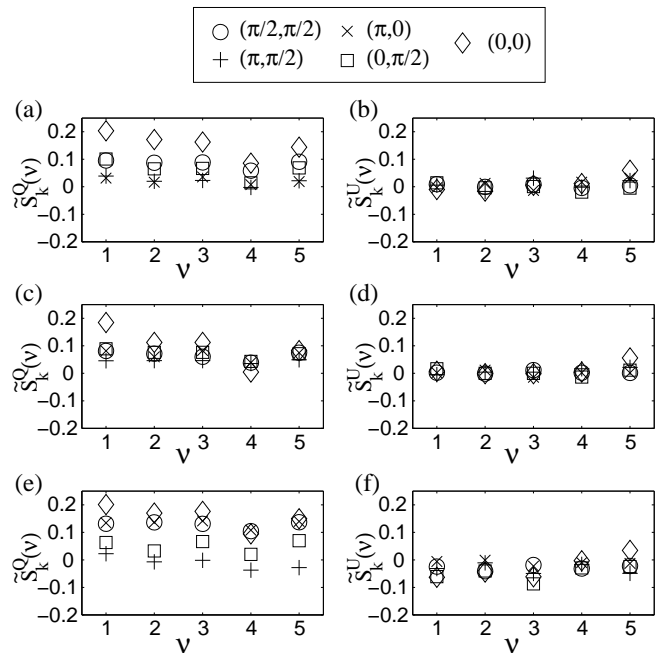}
\caption
{\label{fig:staggered}
$\widetilde{S}_{\bm{k}}^Q(\nu)$ (left panels) 
and $\widetilde{S}_{\bm{k}}^U(\nu)$ (right panels) 
for different momenta $\bm{k}$.
(a) and (b) $t'=-0.3, t''=0.2$. 
(c) and (d) $t',t''=0$.
(e) and (f) $t'=0.3, t''=-0.2$. 
}
\end{figure}

First, consider the average spin density pattern around the hole
\be
S_{\bm{k}}^Y(\bm{i})\equiv\bra{Y_{\bm{k}}} \sum_j S_{j+i}^z 
\widetilde{c}_{j,-1/2} \widetilde{c}_{j,-1/2}\sd \ket{Y_{\bm{k}}} 
\label{eq:spin_density}
\ee
for both $Y=Q$ and $Y=U$.
Fig. \ref{fig:spin} illustrates how different the spin background
is in Q and U states for 
$\bm{k}=(\tfrac{\pi}{2},\tfrac{\pi}{2})$ and $t',t''=0$.
The former preserve an evident staggered pattern while the latter
display an almost uniform distribution of the spin-1/2 introduced
in the system upon doping.
To show that this picture remains valid for other values of $\bm{k}$, 
$t'$ and $t''$, take the average of the staggered magnetization over 
the hole's $\nu^{th}$ NN sites: 
$\widetilde{S}_{\bm{k}}^Y(\nu) \equiv 
- \langle(-)^{i_x+i_y}S_{\bm{k}}^Y(\bm{i})\rangle_{\nu} - 
\tfrac{1}{N-1}\tfrac{1}{2}$
(here $Y=Q,U$). \cite{AVERAGE_S}
Figs. \ref{fig:staggered}(a), \ref{fig:staggered}(c) and 
\ref{fig:staggered}(e) show that for different $\bm{k}$, $t'$ and $t''$
the doped hole in Q states coexists with the staggered 
spin pattern inherited from the undoped system.
This state of affairs is in sharp contrast with the results for
U states, where the AF spin pattern of the undoped system is destroyed 
and the staggered magnetization around the hole 
is close zero and even negative
[Figs. \ref{fig:staggered}(b), \ref{fig:staggered}(d) and 
\ref{fig:staggered}(f)].
One can check that a similar conclusion holds for
$\widetilde{\text{U}}$ states [these are the energy 
eigenfunctions $\ket{\psi_{\bm{k}},J,t',t''}$ when 
$J<J_c({\bm{k}},t',t'')$].

\begin{table}
\begin{ruledtabular}
\begin{tabular*}{\hsize}{c|cc|cc|}
& \multicolumn{2}{c|}{$t'=-0.3; t''=0.2$} &
\multicolumn{2}{c|}{$t'=0.3; t''=-0.2$} \\
$\bm{k}$ & $(\pi,0)$ & $(0,0)$ & $(\tfrac{\pi}{2},\tfrac{\pi}{2})$ & $(0,0)$ \\
$\bm{q}=(0,0)$ & 0.0329 & 0.0343 & 0.0164 & 0.0328 \\
$\bm{q}=(\pi,\pi)$ & 0.5437 & 0.6051 & 0.5293 & 0.5141 \\
\end{tabular*}
\end{ruledtabular}
\caption
{\label{tab:MD} 
$\sum_{\bm{q}}' n_{\bm{k}}^{\widetilde{U}}(\bm{q},+\tfrac{1}{2})\equiv
\sum_{\bm{q}}'\bra{\widetilde{U}_{\bm{k}}}
\widetilde{c}_{-\bm{q},-\tfrac{1}{2}} 
\widetilde{c}_{-\bm{q},-\tfrac{1}{2}}\sd  \ket{\widetilde{U}_{\bm{k}}}$.
$\bm{q}=(0,0)$ results involve sum over $\bm{q}=(0,0)$,
$\bm{q}=(\pm\tfrac{\pi}{2},0)$ and $\bm{q}=(0,\pm\tfrac{\pi}{2})$.
$\bm{q}=(\pi,\pi)$ results involve sum over $\bm{q}=(\pi,\pi)$,
$\bm{q}=(\pm\tfrac{\pi}{2},\pi)$ and $\bm{q}=(\pi,\pm\tfrac{\pi}{2})$.
The model parameters used are relevant to both hole doped cuprates
($J=0.4$, $t'=-0.3$, $t''=0.2$) and electron doped cuprates
($J=0.4$, $t'=0.3$, $t''=-0.2$). \cite{TM0017}
}
\end{table}

\begin{figure}
\includegraphics[width=0.48\textwidth]{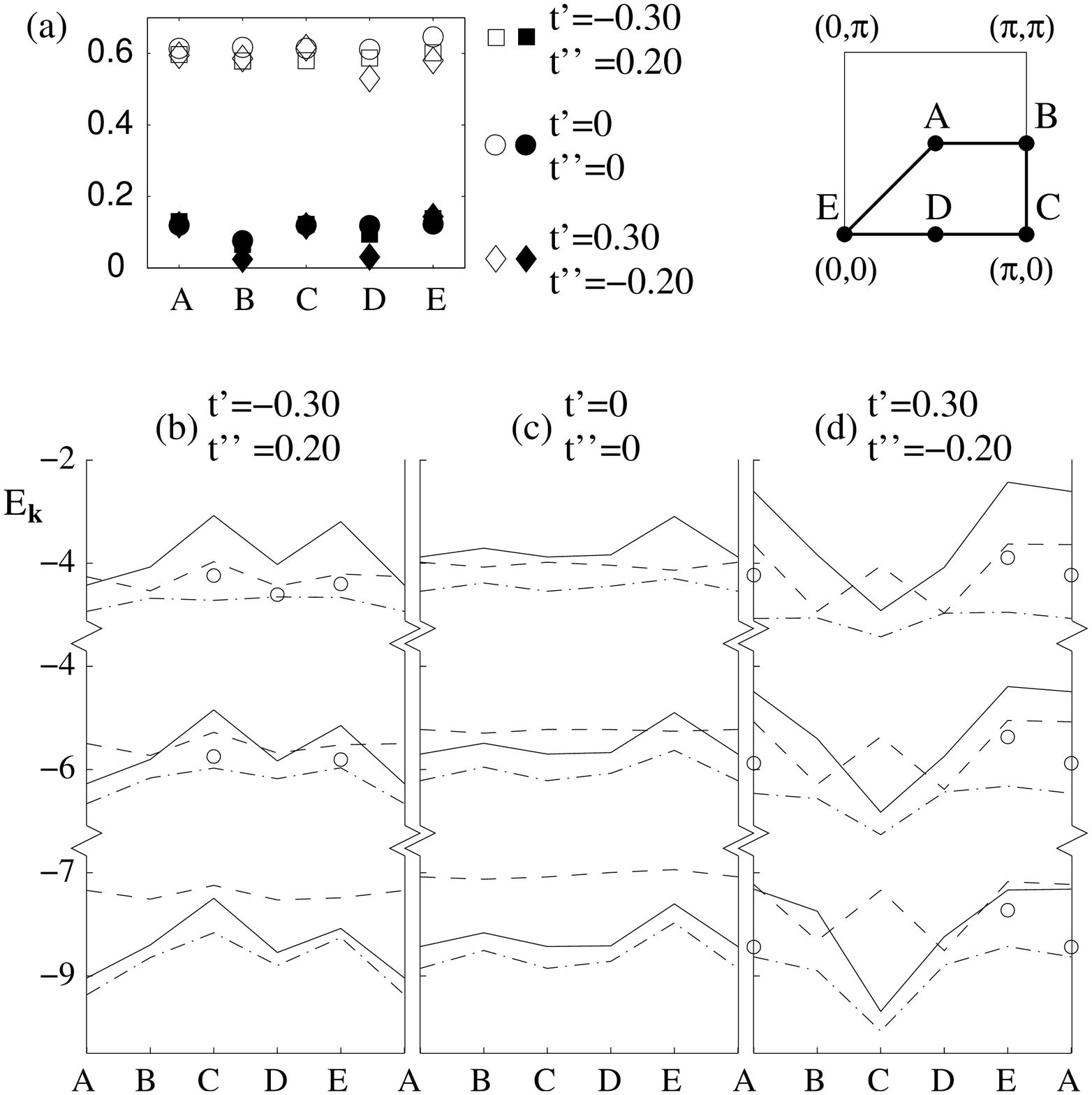}
\caption
{\label{fig:bands}
(a) $\sum_{\bm{q}}' n_{\bm{k}}^U(\bm{q},+\tfrac{1}{2})$.
Empty symbols involve sum over $\bm{q}=(\pi,\pi)$,
$\bm{q}=(\pm\tfrac{\pi}{2},\pi)$ and $\bm{q}=(\pi,\pm\tfrac{\pi}{2})$.
Full symbols involve sum over $\bm{q}=(0,0)$,
$\bm{q}=(\pm\tfrac{\pi}{2},0)$ and $\bm{q}=(0,\pm\tfrac{\pi}{2})$.
(b)-(d) Dispersion relations for $\ket{Q_{\bm{k}}}$
(full line), $\ket{U_{\bm{k}}}$ (dashed line) and $\ket{\psi_{\bm{k}}}$
(dash-dot line). Upper, middle and lower set of dispersions are obtained
for $J$ equal to $0.2$, $0.4$ and $0.7$ respectively.
($\circ$) indicates the best energy obtained by a linear combination of
$\ket{Q_{\bm{k}}}$ and $\ket{U_{\bm{k}}}$ when $J<J_c(\bm{k},t',t'')$ 
(in which case $\ket{\psi_{\bm{k}}} = \ket{\widetilde{U}_{\bm{k}},J,t',t''}$).
}
\end{figure}

In order to complement the above real space picture, 
one also considers the hole momentum distribution function
\be 
n_{\bm{k}}^Y(\bm{q},\sigma)\equiv\bra{Y_{\bm{k}}} 
\widetilde{c}_{-\bm{q},-\sigma} \widetilde{c}_{-\bm{q},-\sigma}\sd  
\ket{Y_{\bm{k}}}
\label{eq:hole_momentum}
\ee
for $Y=Q,U$.
Since Q states bear an electron-like character, the hole momentum 
distribution function
$n_{\bm{k}}^Q(\bm{q},+\tfrac{1}{2})$ is peaked at 
$\bm{q}=\bm{k}$. 
A smaller peak is also observed at $\bm{q}=\bm{k}+(\pi,\pi)$ due 
to the strong AF correlations. \cite{EO9541}
In U states, the hole strongly interacts with the surrounding spins 
and, as a result, the hole momentum distribution function 
$n_{\bm{k}}^U(\bm{q},+\tfrac{1}{2})$ peaks around 
$\bm{q}=(\pi,\pi)$ for all momenta $\bm{k}$ [Fig. \ref{fig:bands} (a)].
Table \ref{tab:MD} illustrates that the hole density in
$\widetilde{\text{U}}$ states also peaks around $(\pi,\pi)$ 
independently of the momentum $\bm{k}$.

The above results confirm that Q states capture the AF 
correlations that persist around the vacancy away from
half-filling.
This is expected since these states have a well defined 
quasiparticle character.
A remarkably different picture holds for the U and 
$\widetilde{\text{U}}$ states, whose quasiparticle spectral weight
vanishes.
As the above spin density results indicate, the spin correlations
in these states spread the extra $S^z=\tfrac{1}{2}$ away from the vacancy.
The resulting loss of spin exchange energy is accompanied by a gain in 
the hole kinetic energy, as it follows from
the hole momentum distribution results which support that, in these
states, the hole always lies around the bare band bottom 
[which is located at $(\pi,\pi)$]. 
This evidence resembles predictions from spin-charge separation scenarios. 
Indeed, within the slave-boson \cite{LN9221,WL9603,LN0617} 
and the doped-carrier frameworks, \cite{RW0501,RW0674}
the electron decays into a charged spinless boson, which condenses at 
$(\pi,\pi)$, and a spin-1/2 chargeless fermion, which 
carries the remaining momentum. 
The above calculations determine the equal time correlations probed by the
quantities in Eq. \eqref{eq:spin_density} and Eq. \eqref{eq:hole_momentum}
in a small lattice 
and, thus, cannot prove the existence (or lack thereof) of true 
spin-charge separation.
Still, they support that, in U and $\widetilde{\text{U}}$ 
states, the lattice spins screen the hole in conformity with 
short-range aspects of spin-charge separation phenomenology.

\subsection{\label{subsec:renorm}Effect on electron dynamics}

The different hopping terms in the $tt't''J$ model Hamiltonian 
[Eq. \eqref{eq:Htj}] move electrons between first, second and third 
NN sites under the no-double-occupancy constraint.
These processes may or may not be restrained by the surrounding spin 
correlations. \cite{PROBE}
For instance, NN hopping is frustrated by the two-sublattice 
structure of AF correlations.
Intra-sublattice hopping processes are, however, consistent
with the staggered pattern of AF correlations which, thus, do
not strongly renormalize $t'$ and $t''$. \cite{RENORM}

\begin{table}
\begin{ruledtabular}
\begin{tabular*}{\hsize}{ccccccc}
$t'$ & $t''$ & $\Delta E^{\psi}$ & $\Delta E^Q$ & $\Delta E^U$ &
$W_{\bm{k'}}^Q$ & $W_{\bm{k''}}^Q$ 
\\
\hline
-0.3 & 0.2 & 0.69 & 1.43 & 0.22 & 0 & 0.75 \\
-0.2 & 0.1 & 0.56 & 0.92 & 0.14 & 0.45 & 0.72 \\
0 & 0 & 0 & 0 & 0 & 0.66 & 0.66 \\
0.2 & -0.1 & -0.75 & -1.08 & -0.10 & 0.76 & 0.50 \\
0.3 & -0.2 & -0.80 & -2.33 & -0.29 & 0.82 & 0 \\
\end{tabular*}
\end{ruledtabular}
\caption{\label{tab:pseudogap}
$\Delta E^Q$, $\Delta E^U$, $\Delta E^{\psi}$ and
$W_{\bm{k}}^Q$ with $\bm{k}=\bm{k'} \equiv (\pi,0)$
and $\bm{k}=\bm{k''} \equiv (\tfrac{\pi}{2},\tfrac{\pi}{2})$ for several 
$t'$ and $t''$ and $J=0.4$.}
\end{table}

The way the spin correlations in U states renormalize $t$, $t'$ and $t''$
is strikingly different though.
Firstly, these correlations are induced as a way to enhance NN hopping.
Secondly, they heavily renormalize $t'$ and $t''$.
To establish the latter fact, consider the hole dispersion in Q states
$E_{\bm{k}}^Q\equiv\bra{Q_{\bm{k}}}H_{tt't''J}\ket{Q_{\bm{k}}}$ and
the hole dispersion in U states
$E_{\bm{k}}^U\equiv\bra{U_{\bm{k}}}H_{tt't''J}\ket{U_{\bm{k}}}$.
Table \ref{tab:pseudogap} displays how the dispersion width between
$(\pi,0)$ and $(\pi/2,\pi/2)$ changes with $t'$ and $t''$  for both 
Q states ($\Delta E^Q\equiv E_{(\pi,0)}^Q-E_{(\pi/2,\pi/2)}^Q$) and U 
states ($\Delta E^U\equiv E_{(\pi,0)}^U-E_{(\pi/2,\pi/2)}^U$). \cite{FLAT}
Indeed, the effect of $t'$ and $t''$ on $\Delta E^Q$ is almost 
one order of magnitude larger than on $\Delta E^U$.

Interestingly, certain spin liquid correlations discussed in the 
context of the $tJ$ model strongly inhibit coherent intra-sublattice 
hopping. \cite{RW0301}
This fact, together with the above results, further supports that 
spin correlations in U states resemble spin liquid correlations at 
short length scales.

\section{\label{sec:anisotropy}Momentum space anisotropy}

In the cuprates' renowned pseudogap metallic regime, 
the low energy physics is determined by the states around the
$(\pi,0)-(0,\pi)$ line. \cite{DS0373}
However, there is a clear distinction between the nodal 
[$\vec{k}=(\pm \pi/2,\pm \pi/2)$] and the antinodal 
[$\vec{k}=(\pi,0),\, (0,\pi)$] regions.
Specifically, ARPES detects an energy difference between the 
single-electron spectral features around $(\pi/2,\pi/2)$ and $(\pi,0)$
(whence the term ``pseudogap'') \cite{DS0373,DY9651} and,
in addition, a strong suppression of the electronic character of 
excitations is observed in the pseudogap region.
These phenomena occur in both hole and electron doped compounds
with a crucial difference:
in the former, low energy quasiparticles appear close to 
$(\pi/2,\pi/2)$ but not around $(\pi,0)$; \cite{ZY0401,YZ0301,RS0301,KF0517}
in electron doped materials, both the pseudogap and the excitations
with little electron-like character are pushed toward the
zone diagonal. \cite{AR0201,KB0322,HP0111} 

This phenomenology is reproduced by the generalized-$tJ$ model, where
it stems from the role of the intra-sublattice hopping parameters 
$t'$ and $t''$. 
Indeed, for $t',t'' = 0$, the quasiparticle states at $(\pi/2,\pi/2)$ 
and $(\pi,0)$ have both comparable energies and spectral weight 
intensities. \cite{MH9117,DN9428}
On the other hand, non-zero $t'$ and $t''$ fit the experimentally 
observed dispersion width along $(\pi,0) - (\pi/2,\pi/2)$.
\cite{GV9466,NV9576,KW9845,T0417}
These intra-sublattice hopping parameters further lead to pseudogap 
states with modified spin background correlations 
\cite{TS0009,LL0301,ME9916} and, thus, with small spectral weight. 
\cite{KW9845,TM0017,T0417,RW0501,KK0614,CC0502} 
As to the difference between the hole and electron doped regimes,
it simply follows from the change in the sign of $t'$ and $t''$. 
\cite{TM9496,GV9466,KW9845,T0417,RW0501,KK0614,CC0502}

The main message of this paper is that the results obtained in 
Sec. \ref{sec:local} and Sec. \ref{sec:properties} provide a 
microscopic two-band picture that rationalizes the above 
generalized-$tJ$ model behavior.
This picture embodies the effect of the Q and U states' short-range 
correlations, which underlies the momentum anisotropic pseudogap 
behavior, as well as its dependence on the electronic density 
(see below).

\subsection{\label{subsec:twoband}Two-band picture}

Sec. \ref{subsec:onehole} identifies the two distinct spin correlations 
that dress the vacancy in low energy single-hole states.
The static properties of these low energy states then follow from 
the reduced two-band Hamiltonian \cite{SINGLEJ}
\be
H_{reduced,\bm{k}} \equiv \left[ \begin{array}{cc}
\bra{Q_{\bm{k}}}H_{tt't''J}\ket{Q_{\bm{k}}} &
\bra{Q_{\bm{k}}}H_{tt't''J}\ket{U_{\bm{k}}} \\
\bra{U_{\bm{k}}}H_{tt't''J}\ket{Q_{\bm{k}}} &
\bra{U_{\bm{k}}}H_{tt't''J}\ket{U_{\bm{k}}}
\end{array}\right]
\label{eq:Hreduced}
\ee
This Hamiltonian yields the two spectral dispersions
observed both by ARPES data \cite{RS0301,KS0318,YZ0301,SR0402}
and by various theoretical studies of the related $tJ$ and Hubbard 
models. \cite{PH9544,MH9545,DZ0016,GE0036,RW0501,KK0614}
It also determines the hybridization of Q and U states and, thus, 
the spectral weight distribution throughout momentum space. 
Therefore, $H_{reduced,\bm{k}}$ must capture the aforementioned
role of $t'$ and $t''$ in the pseudogap phenomenology.

Interestingly, the above role of $t'$ and $t''$ can be discussed 
only in terms of the dispersions $E_{\bm{k}}^Q$ and $E_{\bm{k}}^U$.
To see this, note that $t'$ and $t''$ strongly affect the dispersion 
of a hole surrounded by the AF correlations in Q states 
[in fact, $\Delta E^Q = A (-4t'+8t'')$, 
where the renormalization factor $A\gtrsim 1/2$].
Therefore, $t' < 0$ and $t'' > 0$ increase the energy 
$E_{(\pi,0)}^Q$ and decrease $E_{(\pi/2,\pi/2)}^Q$. 
Consequently, 
$E_{(\pi,0)}^{\psi}\equiv\bra{\psi_{(\pi,0)}}H_{tt't''J}\ket{\psi_{(\pi,0)}}$
also increases and 
$E_{(\pi/2,\pi/2)}^{\psi}\equiv
\bra{\psi_{(\pi/2,\pi/2)}}H_{tt't''J}\ket{\psi_{(\pi/2,\pi/2)}}$
also decreases and, hence, a pseudogap 
$\Delta E^{\psi}\equiv E_{(\pi,0)}^{\psi}-E_{(\pi/2,\pi/2)}^{\psi}$
opens at $(\pi,0)$.
Intra-sublattice hopping is, however, strongly frustrated in U states and,
thus, $t' < 0$ and $t'' > 0$ increase the energy difference 
$E_{(\pi,0)}^Q - E_{(\pi,0)}^U$ while reducing
$E_{(\pi/2,\pi/2)}^Q - E_{(\pi/2,\pi/2)}^U$
[Figs. \ref{fig:bands}(b)-\ref{fig:bands}(c)].
This impacts the extent to which $\ket{Q_{\bm{k}}}$
and $\ket{U_{\bm{k}}}$ hybridize, reducing 
$W_{(\pi,0)}^Q\equiv |\langle \psi_{(\pi,0)}\ket{Q_{(\pi,0)}}|^2$ and
enlarging 
$W_{(\pi/2,\pi/2)}^Q\equiv |\langle \psi_{(\pi/2,\pi/2)}
\ket{Q_{(\pi/2,\pi/2)}}|^2$ (Table \ref{tab:pseudogap}).
For $J=0.4, t'=-0.3, t''= 0.2$ 
the energy $E_{(\pi,0)}^Q$ is so large that the minimum energy obtained 
by a linear combination of $\ket{Q_{(\pi,0)}}$ and $\ket{U_{(\pi,0)}}$ 
becomes higher than that of a different state $\ket{\widetilde{U}_{(\pi,0)}}$
with vanishing quasiparticle spectral weight [Fig. \ref{fig:bands}(b)]. 
As a result, $W_{(\pi,0)}^Q = 0$.
At the same time $W_{(\pi/2,\pi/2)}^Q = 0.75$, so that a sharp difference
is encountered between the electron-like character of nodal and 
antinodal states. \cite{ME9916,TS0009,LL0301}

The change between the cuprates' hole and electron doped regimes
amounts to a change in the sign of $t'$ and $t''$, 
in which case $\Delta E^Q = A (-4t'+8t'')$ changes sign as well.
The above argument then still applies, with the roles of 
momenta $(\pi,0)$ and $(\pi/2,\pi/2)$ interchanged 
[Fig. \ref{fig:bands}(d) and Table \ref{tab:pseudogap}].

\subsection{\label{subsec:doping}Doping dependence}

The above calculation and the ensuing arguments concern a single 
hole surrounded by a spin background and, thus, do not have to
straightforwardly apply in the presence of a finite hole density.
Interestingly, though, a large body of evidence suggests that
single-hole physics is relevant away from half-filling.
Indeed, quantum Monte Carlo, \cite{PH9544,GE0036} exact 
diagonalization \cite{MH9545} and cellular dynamical mean-field 
theory \cite{KK0614} studies show that the two-band structure
identified in the half-filled spectral function below the chemical 
potential remains almost unaffected upon hole doping, whose main 
effect is to transfer spectral weight between the pre-existing 
bands in a momentum dependent manner.
This behavior is expected as long as short-range AF correlations are 
present. \cite{PH9544,MH9545,GE0036}
Since calculations on the $U/t=8$ Hubbard model \cite{DM9582} find 
that such correlations persist around the vacancy up to the 
hole density $x=0.25$, the above two-band picture may 
apply in a wide doping range.
Cuprate ARPES data also displays the two dispersive
features throughout a large portion of the phase diagram,
\cite{RS0301,YZ0301,KS0318,SR0402,DS0373} hence, it
complies with the aforementioned theoretical expectations.

It is well known that the pseudogap phenomenology weakens upon 
increasing the dopant density.
Hence, the pseudogap magnitude diminishes away from half-filling,
\cite{DS0373,TY0403} as does the difference in the electron-like 
character of nodal and antinodal excitations. 
\cite{RS0301,YZ0301,ZY0401,KF0517,AR0201}
This experimental evidence is captured by the naive extension
of the above two-band picture to the finite hole density case.
Indeed, Sec. \ref{subsec:twoband} shows that the momentum
space anisotropic behavior follows from the effect of $t'$ and
$t''$ in the dynamics of holes surrounded by short-range AF 
correlations.
Upon doping, these correlations are gradually replaced by the 
doping induced correlations which prevail in U states.
Since the latter strongly renormalize $t'$ and $t''$, the 
differentiation between the $(\pi/2,\pi/2)$ and $(\pi,0)$ regions
is also gradually depleted.

Refs. \onlinecite{RW0501,RW0674} develop a new mean-field 
approach to the $tt't''J$ model that embodies the
above two-band picture in the presence of a finite hole density.
It explicitly captures the interplay between the mobile holes
and the above two different spin correlations and correctly
describes the microscopic electron dynamics in the 2D doped Mott
insulator.
This assertion is attested by the successful comparison to
other theoretical approaches and especially to a vast portfolio
of non-trivial cuprate ARPES and tunneling conductance data.
The latter include the aforementioned nodal-antinodal dichotomy,
the Fermi arcs, the peak-dip-hump structure, the kink and the extended
flat regions close to $(\pi,0)$ in the electron dispersion and the 
large diversity of tunneling spectra. \cite{RW0501,RW0531}

\section{\label{sec:conclusion}Conclusions}

In this paper, I numerically study how a single mobile hole is 
dressed by the encircling spins within the $tt't''J$ model context. 
Purely local energetic arguments decide whether a staggered
moment configuration or a spin configuration reminiscent
of spin liquid physics prevails around the vacancy.
In the experimentally relevant parameter regime, the competition
between the two spin correlations is very subtle and can be
particularly sensitive to the hole momentum.
Consequently, the electron spectral properties can be extremely 
momentum dependent, displaying a pseudogap and distinct quasiparticle
properties in the nodal and antinodal regions, as observed in 
both hole and electron doped cuprate compounds.
\cite{DS0373,ZY0401,YZ0301,RS0301,KF0517,TY0403,AR0201,KW9845}
AF short-range correlations are gradually depleted upon doping and,
thus, the above differentiation between the nodal and antinodal 
regions is expected to disappear further away from half-filling,
in agreement with the phenomenology of high-T$_c$ superconductors. 
\cite{DS0373,TY0403,ZY0401,KF0517,AR0201}

The above considerations agree with prior work 
\cite{PH9544,KK0614,SP0302,ST0401,RW0531,KK0606}
substantiating that the pseudogap and the resulting momentum space 
anisotropy follow from the local interaction between the doped 
carriers and the short-range spin correlations that strive close 
to the Mott insulating transition.
This paper goes a step further and provides an exact scheme to 
determine the local spin correlations that dress moving carriers.
In particular, it shows that the interplay between the itinerant
doped carrier and the surrounding local moments translates into
the coexistence of two different local correlations, namely the 
staggered moment correlations already present in the undoped system 
and a different type of correlations induced upon carrier doping.
The latter correlations are shown to be responsible for short-range 
phenomenology characteristic of spin-charge separated states and
to have a peculiar impact in the electron dynamics, specifically, 
they strongly renormalize $t'$ and $t''$.

One way to optimize both the hole kinetic energy and the spin 
exchange energy in doped Mott insulators is to spatially 
separate charge and spin degrees of freedom into, say, stripe-like 
configurations. \cite{ZG8991}
The above calculation suggests an alternative scenario:
the quantum superposition of two types of local states which 
separately enhance the $J$ and $t$ terms of the Hamiltonian.
These two states have a drastically different effect on the electron 
dynamics and provide a simple two-band microscopic picture 
of doped Mott insulators.
In this picture, the vacancy is, at times, surrounded by 
staggered moments while, at other instants, spin liquid correlations 
take over in order to facilitate the vacancy's motion.
In the pseudogap momentum space region, the kinetic energy of a 
vacancy surrounded by a local AF spin configuration increases with 
$|t'|$ and $|t''|$, thus tilting the balance in favor of the above 
spin liquid correlations (whose energy is less sensitive to
$t'$ and $t''$).
As a result, in the pseudogap regime, the way the spin background 
dresses the hole at $(\pi/2,\pi/2)$ differs from the way it dresses 
the hole at $(\pi,0)$.
This fact is in consonance with previous numerical evidence for the
approximate decoupling of spin and charge degrees of freedom in
the pseudogap states.
\cite{ME9916,TS0009,LL0301}

Finally, I remark that the above two-band picture provides the basis 
to develop new approximate schemes to describe doped Mott insulators.
The mean-field theory developed in Refs. \onlinecite{RW0501,RW0674}
constitutes one such example.
Remarkably, this approach reproduces a variety of experimental data. 
\cite{RW0501,RW0531}

\begin{acknowledgments}
This work was partially supported by the Funda\c c\~ao 
Calouste Gulbenkian Grant No. 58119 (Portugal), NSF Grant No. DMR-01-23156,
NSF-MRSEC Grant No. DMR-02-13282 and by the DOE Grant No. DE-AC02-05CH11231.
\end{acknowledgments}

\bibliography{hightc}

\begin{thebibliography}{68}
\bibitem[{Uti()}]{Utilde}
\bibinfo{note}{Note the difference between the kets
  $\ket{\widetilde{U}_{\bm{k}},J,t',t''}$ and $\ket{U_{\bm{k}},t',t''}$. The
  former is $J$-\textit{dependent} and is defined for $J < J_c(\bm{k},t',t'')$.
  The latter is $J$-\textit{independent} and is defined for $J >
  J_c(\bm{k},t',t'')$. Similar notation is used for both states since they have
  similar properties in the parameter regime of interest to the cuprates (see
  Sec. \ref{sec:properties})}.
\bibitem[{CHO()}]{CHOICE}
\bibinfo{note}{The above choice of $J=0.2$ and $J=0.6$ could have been made
  different. However, any reasonable set of two values of $J$ in the range
  $[0.2 , 0.8]$ would lead to similar conclusions}.
\bibitem[{WID()}]{WIDERANGE}
\bibinfo{note}{The width of the interval of $J$ values displayed in Table
  \ref{tab:check2d}, namely $[0.2 , 0.8]$, is of the order of the maximum $J$
  value}.
\bibitem[{SIN()}]{SINGLEJ}
\bibinfo{note}{As it follows from the approximate equality in Eq.
  \eqref{eq:decomp}, the states $\ket{Q_{\bm{k}},t',t''}$ and
  $\ket{U_{\bm{k}},t',t''}$ are relatively insensitive to variations of the two
  values of $J$ used to define the above mentioned 2D plane in the single hole
  $tt't''J$ model Hilbert space. In the limit where these two values of $J$
  become infinitesimally close to each other, $\ket{Q_{\bm{k}},t',t''}$ and
  $\ket{U_{\bm{k}},t',t''}$ are constructed from the state vectors
  $\ket{\psi_{\bm{k}},J,t',t''}$ and $\p \ket{\psi_{\bm{k}},J,t',t''}/ \p J$.
  Therefore, one can determine a pair of Q and U states for every single value
  of $J > J_c(\bm{k},t',t'')$}.
\bibitem[{MBZ()}]{MBZB}
\bibinfo{note}{This is the region where the lowest eigenstates lie and whose
  properties one is more concerned about}.
\bibitem[{3D()}]{3D}
\bibinfo{note}{The above formal construction to derive Q and U states can be
  applied, for instance, to the 3D $tt't''J$ model. However, in that case, U
  states are expected to be less relevant since AF correlations, which are more
  robust in 3D, lead to strong quasiparticle properties throughout the
  Brillouin zone}.
\bibitem[{AVE()}]{AVERAGE_S}
\bibinfo{note}{The average magnetization $\tfrac{1}{N-1}\sum_j \langle S_j^z
  \rangle = \tfrac{1}{N-1}\tfrac{1}{2}$, where $N$ is the number of lattice
  sites, is subtracted to reduce finite size effects}.
\bibitem[{PRO()}]{PROBE}
\bibinfo{note}{This is the reason why electronic spectral properties provide an
  indirect probe of the local correlations \cite{SHORTRANGE}}.
\bibitem[{REN()}]{RENORM}
\bibinfo{note}{As shown in Ref. \onlinecite{TS0009}, the hole dispersion in the
  $tt't''J$ model along $(0,\pi)-(\pi,0)$ is controlled by $t' \approx -2t''$
  and is a factor $\sim 2-3$ smaller than expected from the bare parameters.
  This is to be contrasted with the nodal dispersion width which goes from the
  bare value $8t$ down to $2.2J$ due to AF spin correlations.
  \cite{D9463,TS0009,KL8980,MH9117,DS0373} In the parameter regime of interest
  to the cuprates this corresponds to a renormalization by a factor of $\sim
  10$}.
\bibitem[{FLA()}]{FLAT}
\bibinfo{note}{The fact that this width vanishes for $t',t''=0$ is an artifact
  of the $4\times4$ lattice. \cite{DJ9049} In larger systems, the dispersion
  width along $(\pi,0) - (\pi/2,\pi/2)$ for $t',t''=0$ is $\approx 20\%$ of the
  nodal dispersion width \cite{MH9117,DN9428}}.
\end{thebibliography}

\end{document}